\documentclass[sigconf]{acmart}

\AtBeginDocument{%
  }

\setcopyright{acmcopyright}
\copyrightyear{2023}
\acmYear{2023}
\acmDOI{XXXXXXX.XXXXXXX}

\acmConference[EASE 2023]{International Conference on Evaluation and Assessment in Software Engineering}{June 14--16,
  2023}{Oulu, Finland}
\acmPrice{15.00}
\acmISBN{978-1-4503-XXXX-X/18/06}

\usepackage[utf8]{inputenc}
\usepackage{hyperref}
\usepackage{graphicx}
\usepackage{caption}
\usepackage{subcaption}
\usepackage{tabularx}
\usepackage{booktabs}
\usepackage{multirow}
\usepackage{float}
\usepackage{mathtools}
\usepackage{xspace}
\usepackage{soul}
\usepackage{color}
\usepackage{cleveref}
\usepackage{pdfpages}

\newcommand\etal[0]{\emph{et al.}\xspace}
\newcommand{\ie}{\emph{i.e.,}\xspace}
\newcommand{\eg}{\emph{e.g.,}\xspace}
\newcommand{\alp}{\textit{Alphabetical}\xspace}
\newcommand{\rand}{\textit{Random}\xspace}
\newcommand{\diff}{\textit{Code Diff}\xspace}
\newcommand{\mrr}{\emph{MRR}\xspace}
\newcommand{\ndcg}{\emph{nDCG}\xspace}
\newcommand{\preck}{\emph{prec@k}\xspace}

\newcommand{\nrev}{51,566\xspace}

\begin{document}

\title{Assessing the Impact of File Ordering Strategies on Code Review Process}

\author{Farid Bagirov}
\email{farid.bagirov@jetbrains.com}
\affiliation{%
  \institution{JetBrains Research}
  \city{Paphos}
  \country{Cyprus}
}

\author{Pouria Derakhshanfar}
\email{pouria.derakhshanfar@jetbrains.com}
\affiliation{%
  \institution{JetBrains Research}
  \city{Amsterdam}
  \country{The Netherlands}
}

\author{Alexey Kalina}
\email{alexey.kalina@jetbrains.com}
\affiliation{%
  \institution{JetBrains}
  \city{Munich}
  \country{Germany}
}

\author{Elena Kartysheva}
\email{elena.kartysheva@jetbrains.com}
\affiliation{%
  \institution{JetBrains}
  \city{Paphos}
  \country{Cyprus}
}

\author{Vladimir Kovalenko}
\email{vladimir.kovalenko@jetbrains.com}
\affiliation{%
  \institution{JetBrains Research}
  \city{Amsterdam}
  \country{The Netherlands}
}

\begin{abstract}
    Popular modern code review tools (\eg Gerrit and GitHub) sort files in a code review in alphabetical order.
    A prior study (on open-source projects) shows that the changed files' positions in the code review affect the review process. Their results show that files placed lower in the order have less chance of receiving reviewing efforts than the other files. Hence, there is a higher chance of missing defects in these files.
    This paper explores the impact of file order in the code review of the well-known industrial project \href{https://www.jetbrains.com/idea/}{IntelliJ IDEA}.
    First, we verify the results of the prior study on a big proprietary software project.
    Then, we explore an alternative to the default \alp order: ordering changed files according to their code diff.
    Our results confirm the observations of the previous study. We discover that reviewers leave more comments on the files shown higher in the code review.
    Moreover, these results show that, even with the data skewed toward \alp order, ordering changed files according to their code diff performs better than standard \alp order regarding placing problematic files, which needs more reviewing effort, in the code review.
    These results confirm that exploring various ordering strategies for code review needs more exploration.

\end{abstract}

\ccsdesc[500]{Software and its engineering~Empirical software validation}

\keywords{Code Review, Cognitive Bias, Ranking Metrics}


 \maketitle

\section{Introduction}
Code review is one of the common software engineering practices. Its purposes are to identify defects in newly implemented code and improve quality of the software under development~\cite{ackerman1989software,baum2016need,baum2017optimal}.
A recent study by Fregnan \etal~\cite{fregnan2022first} shows that the order of changed files in a code review impacts the quality of the review process and, thereby, indirectly affects the quality of the code. 
They also show that the files ranked higher in the order get more comments compared to the lower ones. 
Moreover, they conducted an experiment with handcrafted examples confirming that bugs in lower files are less likely to be noticed. 
Given these findings, analyzing the ordering strategies and identifying the most beneficial way of arranging changed files in the code review is helpful for developers. 

In this study, we assess different file ordering strategies on \nrev recent code reviews mined from the IntelliJ IDEA project~\cite{intellij}. 
At the first step, we verify results from \cite{fregnan2022first} on this big industrial closed-source project. We analyze the correlation between the location of each changed file in a code review and the number of comments it received.
We also compare the \alp order (which is used by most of the popular code review frameworks \eg \href{https://github.com}{GitHub}, \href{https://www.gerritcodereview.com}{Gerrit}, and \href{https://www.jetbrains.com/space/}{JetBrains Space}) against the \rand order (as the baseline) to confirm the findings of the previous study~\cite{fregnan2022first}.

Additionally, we perform a detailed analysis to compare the \alp order against \diff (as an alternative file ordering strategy). In this experiment, we first compare these two strategies on all code reviews. Then, we take a deeper look and explore the capability of these two strategies on ordering different types of files: (i) Java files, (ii) Kotlin files, and (iii) files that are commented only at the later iterations of a review.

We compare these strategies in terms of their capability to position \textit{problematic} files (\ie files that require more attention and activity from the reviewers) at the top. 
Alike to the previous study~\cite{fregnan2022first}, we consider files that are commented during the review process as the problematic files. 
We use three common ranking metrics to measure the accuracy of file ordering strategies in putting the problematic files higher: \mrr, \preck, and \ndcg.

Similarly to the prior study by Fregnan \etal~\cite{fregnan2022first}, our results show a significant weak correlation between the position of files and the number of comments they receive. 
The results also show that, according to all the ranking metrics, \alp order significantly outperforms the \rand ordering with small effect sizes.
Moreover, according to two ranking metrics (\mrr and \ndcg), \diff is significantly better than \alp ordering strategy in organizing all of the files in the code review. 
Interestingly, this outperformance is more evident when ordering Kotlin files.
In this case, \diff significantly outperforms \alp ordering, according to all ranking metrics, with a medium effect size. 
However, we do not see this considerable difference in ordering Java files and files commented at the later iterations.

The remainder of this paper is structured as follows: Section \ref{sec:methodology} explains the methodology of our experiment. Section \ref{sec:results} presents our results and findings. Finally, Section \ref{sec:conclusion} describes the conclusion and our future works.

\section{Methodology}
\label{sec:methodology}
        This section presents the design of our experiments, the ordering strategies we compare, the code review selection, the metrics we use to compare the ordering strategies, and our data analysis procedure.
    
    \subsection{Experiment Design}
        In our study, we seek to answer the two primary questions:

        \textbf{Question 1. Does the order of files impact the number of comments each file receives in the IntelliJ IDEA project?}
        For this, as was done in~\cite{fregnan2022first}, we calculate Spearman's correlation test (a non-parametric correlation test that shows how well the relationship between two random variables be described as a monotonic function~\cite{myers2013research}) on the number of comments and position of the files.
        Moreover, we compare the default \alp order and the \rand order.
        Since the \alp order is arbitrary with regards to problematic files, comparison with random order can show whether there is a bias in what files get comments.
        To compare both orders, we calculate ranking metrics \mrr, \preck, and \ndcg (see Section \ref{subsec:metrics}) to assess the capability of these strategies in ordering commented files.
        Then, we compare these two ordering strategies via Student's t-test (see Section \ref{subsec:danalysis}).
        
        \textbf{Question 2. Can an order alternative to \alp improve code review process?}
        For this, we analyze the \diff order, which sorts files according to the number of changes in them.
        Similarly to the previous setting, we compare \diff against \alp with ranking metrics and Student's t-test (see Sections \ref{subsec:metrics} and \ref{subsec:danalysis}).
        We also look separately at how orders place problematic Kotlin files, problematic Java files, and problematic files missed during the first review iterations.

    \subsection{File orders}\label{subsec:orders}
        \paragraph{\alp.}
        This is the default order in most code review systems, including one used in JetBrains.
        This ordering strategy positions files based on their full path in alphabetically ascending order.
        
        \textit{\rand.}
        We use this order as a baseline.
        Since the \alp order is rather arbitrary, the comparison with the true random order can shed some light on the bias in the number of comments.
        In the \rand order, files are shuffled randomly.
        
        \textit{\diff.}
        This ordering strategy organizes files according to the number of lines added or removed in them in descending order (\ie files in which more lines are added and removed have a higher priority in this strategy).
        

        \begin{figure}[h]
          \centering
          \includegraphics[width=0.65\linewidth]{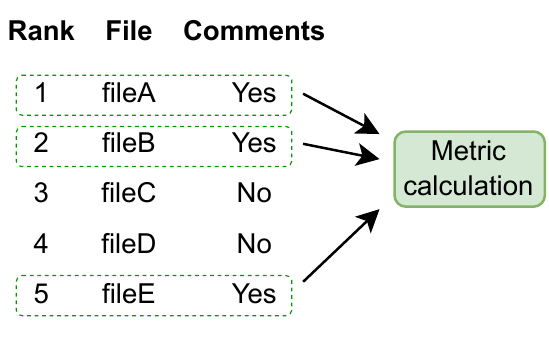}
          \caption{Metric calculation for a single code review.}
          \label{fig:metric}
        \end{figure}


    \subsection{Code Review Selection}
        \label{subsec:data}
        
        Our target project was IntelliJ IDEA~\cite{intellij} --- a large proprietary software product developed by~\href{https://www.jetbrains.com}{JetBrains}.
        For this project, we mined all code reviews from 01.01.2020 to 28.02.2023.
        This resulted in \nrev different code reviews.
        Then, we excluded all code reviews that are insensitive to the file ordering strategy (\ie reviews without comments to the modified files and reviews with only a single changed file).
        For each remaining review, we collect the list of modified files, files that received at least one comment (they are treated as problematic files in our evaluation), and the number of comments they received. 
        We use this data to calculate the ranking metrics using equations presented in Section \ref{subsec:metrics}. Figure \ref{fig:metric} presents an overview of metric calculation using the collected data.

        For the \diff order, we needed to collect file changes.
        However, as code review is an iterative process, the reviewers' comments often cause some changes in files within the code review.
        To correct for this, for each code review, we only consider code changes made before the first comment.
        After that, we remove the code reviews without comments on the changed files or with only one changed file.
        We conducted our experiments on the 9475 resulting reviews.
        Their statisctis are presented in 'Whole data' column in~\Cref{tab:datastats}.

        To compare the \diff and \alp order (Question 2), we additionally look at a specific subset of the problematic files.
        We consider three subsets: Kotlin files, Java files, and files missed in the first review iterations.
        To identify Kotlin and Java files, we look at the file extension (\texttt{.java} for Java, and \texttt{.kt} or \texttt{.kts} for Kotlin).
        We consider files that were missed in the first review (called \emph{missed files} hereafter) as files that changed early but were commented only on the later iterations of the code review.
        To identify them, we collect the time each file is commented and the latest time it is changed during the review process before the first comment.
        Between these two moments, if we see any reviewing and committing activities on any changed file, we consider the file as \emph{initially missed}.
        Then, we filter out all code reviews without problematic files.
        We compare ordering strategies on the remaining reviews according to the positions of the remaining problematic files (\eg count all metrics only on Kotlin files with comments).
        \Cref{tab:datastats} presents the data statistics for Kotlin, Java, and missed files subsamples in the respective columns.

        \begin{table}[]
        \caption{Statistics of the final data used in our study. The 'Whole data' column presents statistics for the whole filtered data.
        The 'Missed files' column contains the statistics of problematic files that are missed in the first iteration of code reviews.
        The 'Java' and 'Kotlin' columns present the statistics of problematic Java and Kotlin files, respectively.}
        \label{tab:datastats}
       \begin{tabular}{l|c|c|c|c}
             & \textbf{\begin{tabular}[c]{@{}c@{}}Whole \\ data\end{tabular}} & \textbf{\begin{tabular}[c]{@{}c@{}}Missed\\  files\end{tabular}} & \textbf{Java} & \textbf{Kotlin} \\ \hline
            \textbf{\begin{tabular}[c]{@{}l@{}}Number of\\ code reviews\end{tabular}}                              & 9475                & 235                   & 5145          & 4075            \\ \hline
            \textbf{\begin{tabular}[c]{@{}l@{}}Average number \\ of commented files\end{tabular}}                  & 1.71                & 1.54                  & 1.47          & 1.58            \\ \hline
            \textbf{\begin{tabular}[c]{@{}l@{}}Average number \\ of modified files\end{tabular}}                   & 14.82               & 21.20                 & 16.26         & 13.91           \\ \hline
            \textbf{\begin{tabular}[c]{@{}l@{}}Average number \\ of added lines \\ per code review\end{tabular}}   & 572.36              & 711.54                & 723.70        & 389.88          \\ 
            \end{tabular}
        \end{table}

    \subsection{Metrics}
    \label{subsec:metrics}
        We use common ranking metrics to compare different orders.
        With our metric selection, we tried to cover different aspects of the ordering strategies.
        We choose \mrr to represent the positions of the first problematic files, \preck to show the number of problematic files appeared at the top, and \ndcg to show full ranking's quality.\
        
        \textbf{MRR.} Reciprocal rank is the inverse position of the first commented file. Mean reciprocal rank, or \mrr, is the average of reciprocal ranks on all code reviews.
    
    
        \textbf{Precision@k (prec@$k$).} For a single code review, precision@k is a portion of commented files in the top $k$. For the whole dataset, precision@k is the average precision on each code review.
        We choose a commonly used value of $k=10$ to represent how order strategies work for the several files on the top.
    
        \textbf{nDCG.} Discounted cumulative gain (DCG) measures the cumulative 'usefulness' of the problematic files based on their position~\cite{jarvelin2002cumulated}.
        If there is a commented file at position $i$, it provides a gain of $\frac{1}{\log{i+1}}$.
        Normalized discounted cumulative gain (nDCG) is DCG that is normalized to the maximum value that DCG can be.
        
        \textbf{Metric calculation.} For \alp and \diff orders, we find the positions of commented files and calculate metrics based on them.
        For the \rand order, we can analytically calculate the expected values of the metrics without sampling actual positions.
        
        Below, we present calculations for metric values done on a single code review.
        $n$ denotes number of modified files, $m$ number of files with comments, and $CF$ is the set of commented files.

        \begin{itemize}
            \item \textbf{MRR}. Let denote $hr$ as the rank of the highest commented file in the order.
            Then, we calculate expectation of MRR in general form: $\sum\limits_{hr=i} MRR(hr=i)p(hr=i)= \sum \frac{p(\mbox{hr=i})}{i}$.
            We estimate $p(hr=i) =  p(hr\leq i) -  p(hr \leq i)$.
            We calculate the latter probability with the help of hypergeometric distribution $HG$ with parameters $n, m, i$ as $p(hr \leq i ) = 1 - HG(x=0)$.
            Hypergeometric distribution with parameters $n, m, i$ is a discrete probability distribution that defines the probability of drawing $x$ objects of a specific type in $i$ draws from $n$ object, where, in total, there are $m$ objects with the desired type.
            In our case, we count the complementary probability of not having any commented files in the first $i$.

            \item \textbf{Prec@k} is calculated with the following equation:
            $$Prec@k= \frac 1{\min(k, n)} \sum\limits_{f\in CF} \mathbf{E}[\mbox{f in top k}] = \\$$$$ \frac 1{\min(k, n)} \sum\limits_{f\in CF} \frac{\min(k, n)}{n} = \frac{m}{n} $$

            \item Normalisation in \textbf{nDCG} is not affected by randomness, and only $DCG$ should be calculated by the following equation:
            $$DCG = \sum\limits_{f\in CF} \mathbf{E}\frac{1}{\mbox{position of $f$ + 1}} = \frac{m}{n} \sum\limits_{i=1}^{n} \frac{1}{i + 1}$$
        \end{itemize}

        \subsection{Data Analysis}
        \label{subsec:danalysis}

        \paragraph{Metric comparison (To answer both \textbf{Question 1} and \textbf{Question~2})} All the metrics we use in this study are averaged over all code reviews.
        Therefore, we can consider them as an expected value over some code review distribution.
        For comparison, we calculate each ranking metric for each code review and use paired Student's t-test, with $\alpha = 0.05$ for Type I errors, to assess the significance of the results ($p-value \leq 0.05$ indicates the significance of our results).
        We use Student's t-test for two related samples since we need to compare the mean values of two samples. 
        Moreover, since the number of samples is high due to CLT mean, values of the metrics approximately have a normal distribution~\cite{goldberger1964econometric}.
        We use Cohen's d test, which is standard for t-test~\cite{cohen2013statistical}, to calculate effect sizes
        (we treat effect sizes < $0.2$ as small and from $0.2$ to $0.5$ as medium).

        \textit{Correlation test (To answer \textbf{Question 1})}. 
        To confirm the correlation between the file's position and the number of comments it gets, we perform the Spearman correlation test~\cite{myers2013research}.
        We perform this correlation test on a normalized number of comments and file positions.
        We normalize the number of comments to a file to the total number of comments in the code review, and the file position is normalized to the total number of files in the review.
        With normalization, we remove the effect of reviews of different sizes (\ie positions with high numbers have fewer comments since they can appear only in a small portion of code reviews).
\section{Findings and Results}
\label{sec:results}
\subsection{Question 1.}
\paragraph{Correlation between position and number of comments} 
We obtained p-value $<0.001$ and $\rho=-0.121$, which aligns with the results of Fregnan \etal~\cite{fregnan2022first}: there is a correlation between the number of comments and position, but a rather small one.

\begin{table}[h]
\caption{Comparison of \alp order and \rand order on the whole data. Bold highlights values that are significantly higher than their counterpart (p-value < $0.05$).}
\label{tab:comp_rand}
\begin{tabular}{l|c|c|c}
& \textbf{MRR}     & \textbf{prec@k}  & \textbf{nDCG}    \\ \hline
\textbf{\alp order} &  \textbf{0.537 }          & \textbf{0.285}            & \textbf{0.646}            \\ \hline
\textbf{\rand order}       &  0.491           & 0.278            & 0.607            \\ \hline
\textbf{effect size}        & 0.158             & 0.145             & 0.187             \\
\end{tabular}
\end{table}

\begin{table}[h]
\caption{Comparison of \alp order and \diff order on the whole data. 
Bold highlights values significantly higher than their counterpart (p-value < $0.05$ ).
}
\label{tab:comp_diff}
\begin{tabular}{l|c|c|c}
& \textbf{MRR}   & \textbf{prec@k} & \textbf{nDCG}  \\ \hline
\textbf{\alp order} & 0.538          & \textbf{0.285}  & 0.646          \\ \hline
\textbf{\diff order}         & \textbf{0.560} & 0.283           & \textbf{0.657} \\ \hline
\textbf{effect size}        & 0.06           & 0.04            & 0.04           \\ 
\end{tabular}
\end{table}

\textit{\alp order vs \rand order} 
We present the results of this comparison in \Cref{tab:comp_rand}.
In the rows ``\alp order'' and ``\rand order'', we list metric values for the \alp and \rand orders, respectively.
Metric value in bold highlights the significantly better order strategy (\ie p-value $<0.05$).
This table also reports p-values and effect sizes of point-wise metric comparisons in the respective rows.
The results show that the \alp ordering places the commented files significantly higher, but the effect size is small ($<0.2$).
Since both orders are arbitrary concerning actual problematic files, we can conclude that there is a small bias towards the \alp order.
This result aligns with the results from the correlation test and also indicates that the files that are higher tend to get more comments.

\subsection{Question 2.}
\paragraph{\alp order vs \diff order}
\Cref{tab:comp_diff} shows the comnparison between these two ordering strategies. 
Despite the bias toward the \alp order, \diff significantly outperforms \alp in \mrr and \ndcg.
This can be interpreted as \diff order being better in showing the first problematic file higher (\mrr) and has a better ranking as a whole (\ndcg).
However, despite the difference, the values of effect sizes are low ($<0.2$).
This can be caused by heavy bias from the \alp order or by low sensitivity of the selected metrics.
Verification of these hypotheses is our plan for the future work.

Additionally, we compared the \alp order and the \diff order of different types of \emph{problematic} files (Tables \ref{tab:comp_diff_missed}, \ref{tab:comp_diff_java}, and \ref{tab:comp_diff_kotlin}).
Mostly, there is no significant difference between \alp and \diff order.
However, the \diff order positions Kotlin files higher with the relatively larger effect size ($>0.2$).
It is worth noting, that results reported here can underestimate real effect, since historical data were sorted in \alp order.

\begin{table}[]
\caption{Comparison of \alp order and \diff order on problematic file that were missed in the first iteration of code reviews.
}
\label{tab:comp_diff_missed}
\begin{tabular}{l|c|c|c}
& \textbf{MRR} & \textbf{prec@k} & \textbf{nDCG} \\ \hline
\textbf{\alp order} & 0.364        & 0.175           & 0.505         \\ \hline
\textbf{\diff order}         & 0.370        & 0.174           & 0.516         \\ \hline
\textbf{effect size}        & 0.01         & 0.02            & 0.04          \\
\end{tabular}
\end{table}

\begin{table}[]
\caption{Comparison of \alp order and \diff order on Kotlin problematic files.
}
\label{tab:comp_diff_kotlin}

\begin{tabular}{l|c|c|c}
& \textbf{MRR} & \textbf{prec@k} & \textbf{nDCG} \\ \hline
\textbf{\alp order} &   0.448      & 0.251         & 0.583         \\ \hline
\textbf{\diff order}         & \textbf{0.538}        & \textbf{0.253}           & \textbf{0.639}         \\ \hline
\textbf{effect size}        & 0.23         & 0.03            & 0.21          \\ 
\end{tabular}
\end{table}

\begin{table}[]
\caption{Comparison of \alp order and \diff order on Java problematic files. 
}
\label{tab:comp_diff_java}
\begin{tabular}{l|c|c|c}
& \textbf{MRR} & \textbf{prec@k} & \textbf{nDCG} \\ \hline
\textbf{\alp order} &   0.534      & \textbf{0.264}         & 0.644         \\ \hline
\textbf{\diff order}         & 0.544        & 0.261           & 0.645         \\ \hline
\textbf{effect size}        & 0.02         & 0.05            & 0.01          \\ 
\end{tabular}
\end{table}


\section{Conclusion and future work}
\label{sec:conclusion}
The order of files in code review impacts the level of attention they receive from reviewers. 
Hence, using a file ordering strategy that places the files likely to require more reviewing effort on top of the list is potentially beneficial. 

In this work, we have applied three different file ordering strategies (\alp, \rand, and \diff ordering) on \nrev code reviews mined from the IntelliJ project. We have evaluated these ordering techniques in terms of their capability of placing problematic files at the top of the order. 
For this evaluation, we have used three ranking metrics: \mrr, \ndcg, and \preck. Our results confirm the correlation between the positions of changed files in the code review and the number of comments they receive. We also show that an alternative ordering technique (\diff ordering) outperforms the standard \alp ordering. 
 
Our results suggest that evaluating and identifying a file ordering strategy better than standard \alp order requires more exploration.
Hence, in our future work, we aim to compare ordering in the unbiased setting and explore other promising ordering strategies (\eg using large language models to identify problematic files). Additionally, we plan to test promising ordering strategies in actual workflows and evaluate their effect on the software quality.

\bibliographystyle{ACM-Reference-Format}
\bibliography{main}

\end{document}